\documentclass[]{spie}  

\usepackage{amsmath,amsfonts,amssymb}
\usepackage{graphicx}
\usepackage[colorlinks=true, allcolors=blue]{hyperref}

\title{Planar Silicon Metamaterial Lenslet Arrays for Millimeter-wavelength Imaging}

\author[a]{Christopher M. McKenney}
\author[b]{Jason E. Austermann}
\author[b]{James A. Beall}
\author[a]{Nils W. Halverson}
\author[b]{Johannes Hubmayr}
\author[a]{Gregory Jaehnig}
\author[c]{Giampaolo Pisano}
\author[a]{Sarah A. Stevenson}
\author[d]{Aritoki Suzuki}
\author[e]{Jonathan A. Thompson}

\affil[a]{Center for Astrophysics and Space Astronomy, University of Colorado, Boulder, CO, USA}
\affil[b]{NIST Quantum Sensors Group, Boulder, CO, USA}
\affil[c]{Dipartimento di Fisica, Sapienza Università di Roma, Piazzale Aldo Moro 2, 00185 Roma, Italy}
\affil[d]{Lawrence Berkeley National Laboratory, Berkeley, CA, USA}
\affil[e]{School of Physics and Astronomy, Cardiff University, The Parade, Cardiff UK, CF24 3AA}
\authorinfo{Further author information: (Send correspondence to C.M.M)\\C.M.M.: E-mail: christopher.m.mckenney@colorado.edu\\  N.W.H.: E-mail: nils.halverson@colorado.edu}

\begin{document} 
\maketitle

\begin{abstract}
Large imaging arrays of detectors at millimeter and submillimeter wavelengths have applications that include measurements of the faint polarization signal in the Cosmic Microwave Background (CMB), and submillimeter astrophysics. We are developing planar lenslet arrays for millimeter-wavelength imaging using metamaterials microlithically fabricated using silicon wafers. This metamaterial technology has many potential advantages compared to conventional hemispherical lenslet arrays, including high precision and homogeneity, planar integrated anti-reflection layers, and a coefficient of thermal expansion matched to the silicon detector wafer. Here we describe the design process for a gradient-index (GRIN) metamaterial lenslet using metal-mesh patterned on silicon and a combination of metal-mesh and etched-hole metamaterial anti-reflection layers. We optimize the design using a bulk-material model to rapidly simulate and iterate on the lenslet design. We fabricated prototype GRIN metamaterial lenslet array and mounted it on a Polarbear/Simons Array 90/150~GHz band transition edge sensor (TES) bolometer detector array with sinuous planar antennas. Beam measurements of a prototype lenslet array agree reasonably well with the model simulations. We plan to further optimize the design and combine it with a broadband anti-reflection coating to achieve operation over 70--350~GHz bandwidth. 

\end{abstract}


\section{INTRODUCTION}
\label{sec:intro}  

Astrophysical discoveries at millimeter and submillimeter wavelengths are being driven by recent advancements in imaging array technologies. For example, recent advances in mm-wave measurements of the Cosmic Microwave Background (CMB) have been enabled by large $\sim10,000$ pixel focal plane arrays spread across multiple telescopes, with even higher pixel counts planned for the near future. In addition, the application space for millimeter and submillimeter polarimetry, imaging, and spectroscopy is large, and future discoveries are sure to be enabled by further technology advancements. 

The efficient coupling of radiation from telescope optics to large-scale arrays of broadband detectors presents important challenges for current and future cosmology and astronomy missions. Design considerations include: manipulating the beam size to control spillover and optical loading on the detectors, controlling beam symmetry to reduce systematics, optimizing the focal plane filling factor to improve sensitivity, and ease of fabrication, to name a few. In addition, the optical coupling technology must be mechanically compatible with detector arrays monolithically fabricated on silicon that operate at sub-Kelvin temperatures. 

Hemispherical lenslets coupled to sinuous antennas are being widely used on ground-based CMB experiments including Polarbear/Simons Array\cite{arnold12,siritanasak16}, SPT-3G\cite{anderson18}, Simons Observatory\cite{galitzki18}, and are proposed for future CMB space missions, including LiteBIRD\cite{sekimoto18} and the Probe of Inflation and Cosmic Origins (PICO) \cite{hanany19}, a NASA Probe-class mission concept study for the 2020 decadal panel. Compared to other optical coupling technologies, lenslet-coupled arrays are inherently broadband, thus making efficient use of the focal plane with multi-chroic detectors. However, they are challenging to fabricate and, unlike feedhorns, must be anti-reflection (AR) coated to achieve the required optical efficiency.  Current anti-reflection coatings are generally made of molded epoxies \cite{siritanasak16} or molded and glued PTFE\cite{nadolski18}, both of which have issues with adhesion, non-uniformity, and mechanical failures during cool down.  


We are developing  planar lenslet arrays for mm-wavelength imaging arrays using metamaterials monolithically fabricated on silicon wafers. Instead of curved optical surfaces, the lenslets consist of a stack of silicon wafers each patterned with a periodic array of sub-wavelength metal-mesh features. Beam-forming is accomplished by creating a metamaterial with a radial or axial gradient in the effective index of refraction (called a GRIN). This metamaterial technology has many potential advantages compared to conventional hemispherical lenslet arrays. The beam-forming elements are flat, lending themselves to an integrated broadband planar anti-reflection layer. Since they are micro-fabricated on monolithic silicon wafers, the metamaterial lenslet arrays are precisely toleranced, homogeneous, have a coefficient of thermal expansion matched to the silicon detector wafer, and can be aligned to the detector wafer using lithographically etched features. In addition, the metamaterial lenslet arrays have a high focal plane filling factor and are scalable to high frequencies, thus overcoming many of the inherent limitations of curved lenslet arrays. 

Work to date has demonstrated that metamaterial lenslets can be fabricated using standard lithographic techniques, and that our measurements are largely in agreement with simulations, giving confidence in the technology and our methods. In this paper, we first give an overview of the metamaterial GRIN lenslet principles of operation and design methodology, including the development of a bulk-model equivalent of the metamaterial which allows rapid iteration and optimization of the design. We then describe the design and preliminary measurements of a prototype 19-element lenslet array which we tested on a Polarbear/Simons Array (PB/SA) ``PB2a'' detector array designed for operation at 90 and 150 GHz. We finish with future work, including new lenslet designs which should significantly improve optical properties, and implementation of broadband AR coatings.



\section{Planar Metamaterial GRIN Lenslet Principles}
\label{GRINprinciples}

In a lenslet-coupled planar antenna element, the lenslet collimates the spherical-wavefront Gaussian beam launched by the antenna thereby enlarging the Gaussian beam waist. This serves to more efficiently couple the radiation to the telescope optics, allowing pixels to be close-packed while
leaving room on the device wafer between antennas for detectors,
band-defining filters, and bias/readout wiring. Metamaterials, materials with sub-wavelength features used to create desired electromagnetic properties, can be used to create a lens with a flat substrate. This can be accomplished by creating a metamaterial with a radial or axial gradient in the effective index of refraction (called a GRIN). The collimated wavefront can then be easily coupled to planar anti-reflection (AR) structures which makes broadband operation possible.

\begin{centering}
\begin{figure}[h!]
\includegraphics{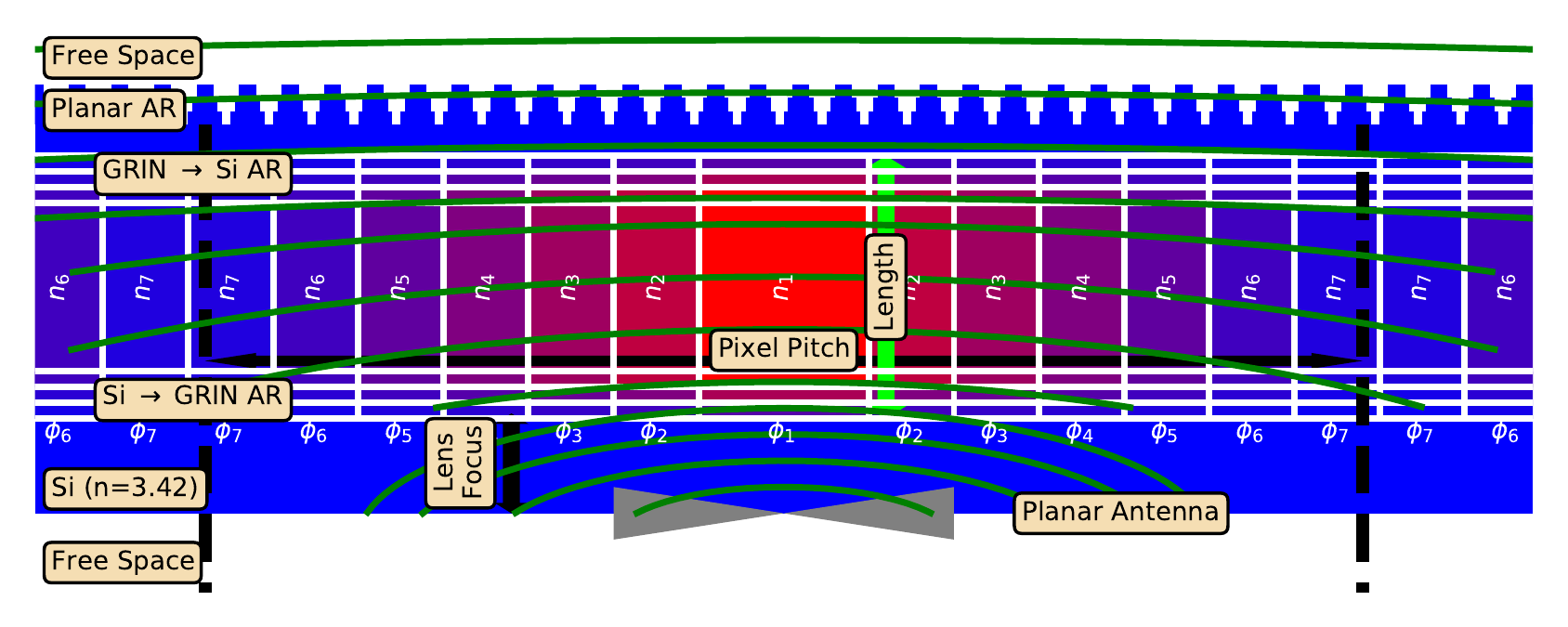}
\caption{Principles of a GRIN coupled lenslet. In a time reverse model, incident radiation (green) from the planar antenna is intercepted by the GRIN material with a radially varying index.  The GRIN is subdivided in to regions with decreasing effective index ($n_{1}$ to $n_{7}$), which then repeats in neighboring pixels.  The overall length of the lens and effective indices are chosen to align the wavefront at the exit of the lens.  Between the GRIN and bulk silicon is an AR layer which has meshes that step down in size between the two.  An additional AR unit, shown here as tapered silicon, serves as an interface between the silicon and free space.}
\label{fig:GRINBasics}
\end{figure}
\end{centering}

A basic schematic showing the cross-section of the operation of the GRIN lenslet is shown in Fig. \ref{fig:GRINBasics}.  In this design, a beam is shown emanating from a planar antenna on the bottom of a silicon detector wafer. A GRIN lenslet is mounted above the detector wafer with a radial gradient from a high index of refraction metamaterial at center ($n_{1}$), to a low index at the outer edge of the lenslet ($n_{7}$). In our design, metal mesh structures embedded in multiple layers of silicon are used to create a high ($n_{1} \sim 4.5$) effective index in the lenslet center, which tapers to the index of silicon ($n_{7} \sim n_{Si} \approx 3.4$) at the lenslet edge. The thickness in silicon from the planar antenna to the bottom of the GRIN lenslet and the lenslet focal length can be optimized to obtain the the desired Gaussian beam waist size at the top of the lenslet. In addition, AR coatings must be implemented between the silicon detector/spacer wafer and the metamaterial lenslet, and at the top from the metamaterial lenslet to free space. 

In the time-reverse sense, the behavior of the sinuous lenslet can be approximated as launching a Gaussian beam with size of the beam waist varying proportionally to wavelength.  This beam intercepts the lens, with the phase at the center of each index-region labeled in Fig. \ref{fig:GRINBasics} as $\phi_{i}$ with $i=1-7$.  While field interactions will affect this beam, some approximation of the length of the GRIN lenslet can be calculated using the phase difference between the center and edge for a given lens focus and the relative indices.  The length of the lens ($L$) will add additional effective path length to align both phases at the lens exit,
\begin{equation}
    L = \left(\frac{\lambda_{0}}{2 \pi}\right) \frac{\phi_{7} - \phi_{1}}{n_{1} - n_{2}}
\label{eq:length}
\end{equation}
where $\lambda_{0}$ is the free space wavelength.  Once the length is fixed, the approximate values for intermediate indices can be determined by solving Eq.~\ref{eq:length} using the phase values and fixing $L$.

As an example for scale, the sinuous antenna used in the PB/SA PB2a detector arrays has a relative phase difference between center and edge at a distance of 1.1 mm from the antenna of $\sim 4\pi$ at 90 GHz, and the metamaterials explored had effective indices spanning bulk silicon ($n_{i} \approx 3.4$) to $n_{i} \approx 4.5$, suggesting optical path lengths of $L \sim$ 3 mm.  Ultimately FEM simulations fully accounting for near field and bending of the rays are needed to fine tune parameters.

Antireflection (AR) layers must be implemented at both the bottom and top of the GRIN metamaterial lenslet. AR layers between silicon and the higher metamaterial index of refraction can be implemented using the same metamaterial as is used in the GRIN, as shown separated by the horizontal white lines n Fig. \ref{fig:GRINBasics}. In addition, a planar silicon to free space tapered AR structure can be integrated at the top of the lenslet, schematically depicted by tapered blue columns in Fig. \ref{fig:GRINBasics}. Unlike conventional hemispherical lenslets, the AR coating does not reduce the usable lens width, allowing for a larger waist than is possible in a hemispherical lenslet and thereby improving the optical efficiency.

\clearpage

\section{Metamaterial Implementation and Modeling}
\label{sec:construction}

To realize a lenslet as shown in Fig. \ref{fig:GRINBasics}, metamaterials are needed which simultaneously satisfy several conditions.  They must be easily scalable, mechanically robust, CTE  (coefficient of thermal expansion) matched to silicon in cryogenic environments, and have optical features which work within these constrained focal plane geometries.  It must also be possible to easily model lenslet designs to allow their optimization.

We have modeled, built and tested a variety of metamaterial options made from lithographically defined structures on silicon substrates, which can then be stacked (see also Pisano et al 2020 \cite{pisano20}).  Deep Reaction Ion Etching (DRIE) technology allows for mating structures to be lithographically defined and layer-to-layer alignments of $\sim 10 \; \mu\text{m}$, or better, are feasible.  We have found that different types of metamaterial are appropriate for different functions in the lenslet design: an embedded metal mesh for the lens body and deep etched holes for silicon to vacuum matching.

\subsection{Modeling Challenges}

Simulation of metamaterial options was carried out using Ansys HFSS with a detailed model of the metamaterial construction.  However, it was discovered that the memory required to sufficiently model a single metamaterial element was significantly larger than anticipated.  For instance, a $100\times100\times100 \; \mu\text{m}^{3}$ silicon cube with a single metal mesh could require more than 50 MB to sufficiently converge the optical properties, which is acceptable for investigating the properties of these metamaterials in an isolated manner.  However, when assembled into a complete lenslet requiring $\sim 10^{5}$ elements or more, this level of model detail leads to memory requirements of several terabytes and long convergence times, even on a cluster computer.  This makes the iterative optimization of designs infeasible. 

We therefore developed a bulk-model equivalent of the metamaterial in which each element is replaced with a uniform block with anisotropically defined permittivity and permeability.  To calculate these parameters, a stack of metamaterials is simulated using boundary conditions which place the stack in an infinite periodic array and excite it with a plane wave from each direction.  By changing the number of elements, essentially the length of the stack, the calculated scattering parameters are used to extract the phase velocity.  Each metamaterial option has effective components in the z direction ($\epsilon_{eff,z}, \; \mu_{eff,z}$) as well as in the x,y directions, which because of symmetry are referred hereafter to as the co-planar components ($\epsilon_{eff,Co}, \; \mu_{eff,Co}$).  For consistent labeling, illumination of the stack by a plane wave traveling in the z-direction is referred to as the \textit{Co-Planar} excitation.  When the plane wave travels along the x or y plane, tangent to the antenna and lenslet, it is referred to as \textit{Normal E} or \textit{Normal H} excitation, with the difference being the polarization of the E and H fields.

The effective bulk permittivity and permeability are related to the phase velocity measured for each illumination orientation by: 
\begin{subequations}
\begin{align}
&v_{ph,Co} = c / \sqrt{\epsilon_{eff,Co} \; \mu_{eff,Co}} \\
&v_{ph,NH} = c / \sqrt{\epsilon_{eff,Co} \; \mu_{eff,z}} \\
&v_{ph,NE} = c / \sqrt{\epsilon_{eff,z} \; \mu_{eff,Co}} 
\end{align}
\label{Eq:phasevel}
\end{subequations}
where $c$ is the speed of light in free space and the subscripts $Co$, $NH$ and $NE$ refer to the three excitations used.  While this presents three equations and four unknowns, there are straightforward physical arguments which reduce Eq. \ref{Eq:phasevel} to a solvable number of optical parameters for each type of metamaterial.

\subsection{Metal Mesh Embedded Silicon}

Metal squares patterned on silicon add effective capacitance which can be modeled primarily as increased permittivity for propagation normal to the grids.  This leads to a material with a higher effective index that can be easily tailored by altering both the pitch and size of the squares.   This is ideal for backside coupling to silicon based detectors, as the higher index avoids problems from total internal reflection which are present in a lower-index coupled lens.

\begin{figure}[ht!]
\includegraphics{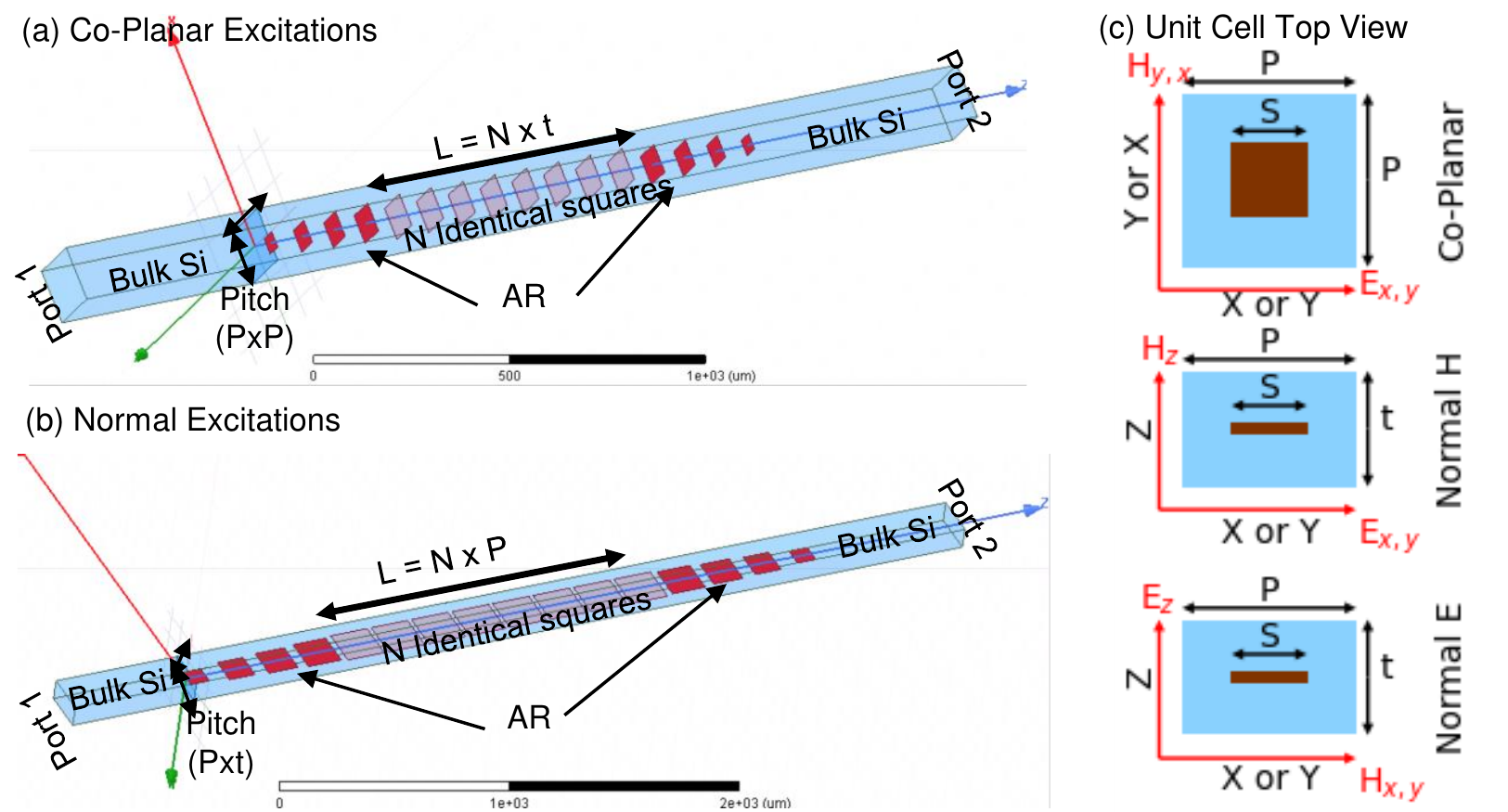}
\caption{Overview of the metal-mesh metamaterial geometry used in simulations to extract equivalent bulk optical parameters. These represent unit cells simulated in a periodic array.  (a) Co-Planar excitation.  In this geometry the plane wave travels normal to the antenna and lenslet and the E and H fields are polarized along the (x,y) axis.  This is the dominant mode in the lenslet.  The length is varied by adding more squares with a spacing given by the substrate thickness, t.  (b) Normal excitations.  In this mode, the plane wave is traveling tangent to the antenna.  The E and H fields can be polarized in two different geometries at the ports.  The length is varied by adding additional layers separated by one lithographic pitch, P.  (c) Top view of the 3 types of excitation of the metamaterial structures showing the E and H fields for each.  The thickness of the metal mesh elements in the Normal excitations is not to scale for these.}
\label{fig:MetalMeshTop}
\end{figure}

Schematics of the structures  simulated to determine equivalent properties are shown in Fig. \ref{fig:MetalMeshTop}.  A single structure is used to investigate the Co-Planar optical properties and one additional for the normal optical components.  For the normal excitations, the E and H field polarizations are simply reversed, allowing the same structure to be re-used with only modified port excitations.

Embedded between each port and the metal mesh grids are additional grids with square side lengths that taper in size from the bulk silicon to the designed side length.  These act as anti-reflection structures in the bulk property simulations and keep the reflected power, $S_{11}$, less than -30 dB. 

\begin{figure}[ht!]
\includegraphics{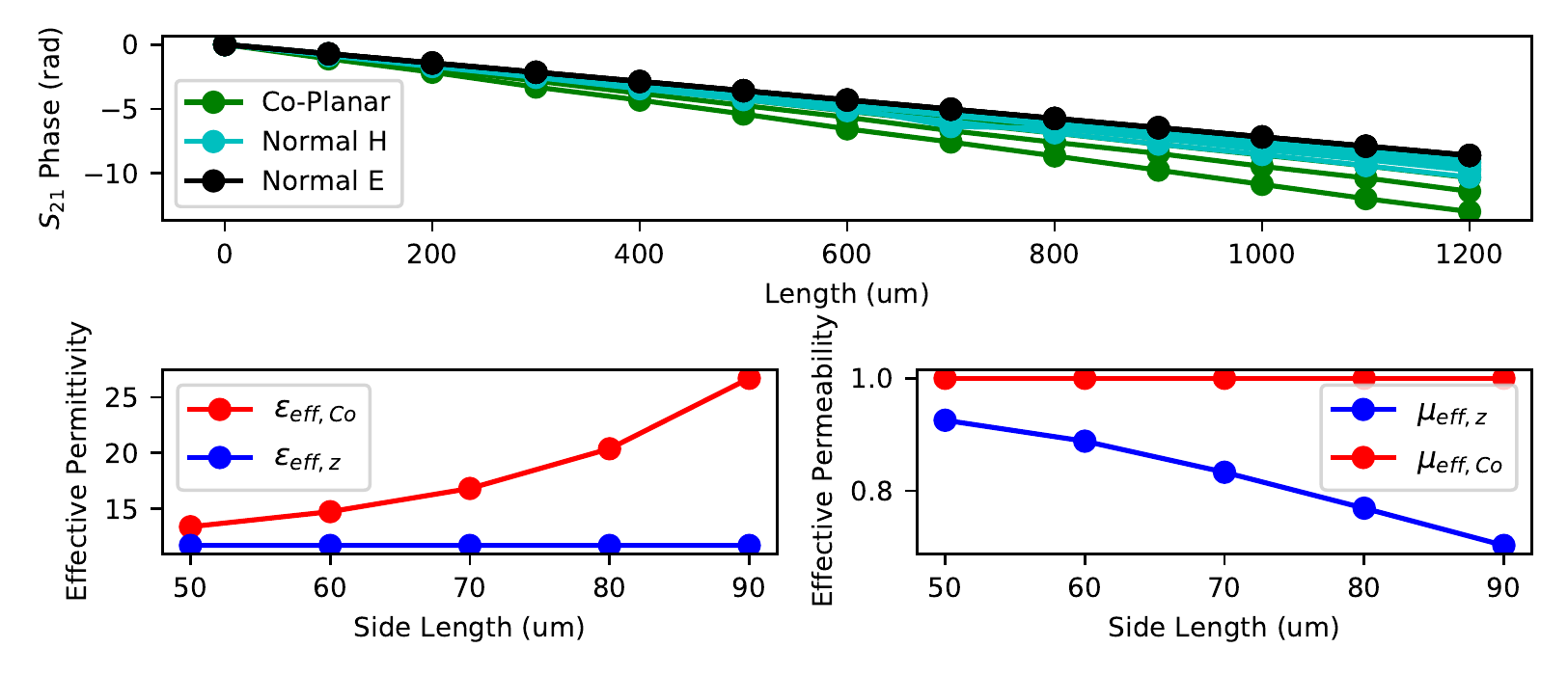}
\caption{Overview of simulation results determining optical properties along different optical excitations.  (a) Phase change as function of length.  Phases are plotted for a series of different metal mesh widths using each excitation.  (b) Results of equivalent parameters.  For the effective permittivity, fitting the data shows an increase with side length for the Co-Planar excitation as expected, whereas fitting the normal direction recovers the bulk value of silicon (11.7).  For permeability, a similar phenomenon is shown where the permeability decreases with side length in the z direction but equals free space when fit with the Co-Planar data.}
\label{fig:MetalMeshAllRots}
\end{figure}

For the metal mesh structure, two of the optical properties in Eq. \ref{Eq:phasevel} can be deduced from physical arguments.  Any electric field polarized in the z direction will not be affected by the capacitance of the metal squares, and therefore $\epsilon_{eff,z} = \epsilon_{eff,Si}$, i.e. the bulk substrate value.  Further, any magnetic components in the (x,y) plane cannot excite currents so long as the mesh elements are significantly thinner than the skin depth, and therefore $\mu_{eff,Co} = 1$.  In the co-planar excitation the phase velocity will only be impacted by $\epsilon_{eff,Co}$, but when the H-field is polarized normal to the metal mesh, induced currents will cause a suppressed value for the effective normal permeability, $\mu_{eff,z}$.  Simulations carried out with realistic copper resistivities and thickness ($\sim 400 \; \text{nm}$) found less than 0.1\% deviation on simulated phase velocities even at 300 GHz, and therefore in the interest of simulation time and memory the squares are modeled in HFSS as perfect conductors.  

The phase velocities are calculated for any particular side length and the relevant optical parameters solved according to the reduced equations,
\begin{subequations}
\label{eq:metalmeshvph}
\begin{align}
&v_{ph,Co} = c / \sqrt{\epsilon_{eff,Co} } \\
&v_{ph,NH} = c / \sqrt{\epsilon_{eff,Co} \mu_{eff,z}} \\
&v_{ph,NE} = c / \sqrt{\epsilon_{Si} } 
\end{align}
\end{subequations}
which results in three equations and three unknowns.  Simulation data show agreement between the assumption in \ref{eq:metalmeshvph}(c) and the defined bulk value of the permittivity of silicon as well as the assumption that for sufficiently thin films, $\mu_{eff,Co} = 1$.

Fig.~\ref{fig:MetalMeshAllRots} presents an overview of extracted bulk-equivalent optical properties as a function of metal mesh square side length.  
As expected the effective index generally increases as the side length (fill fraction) of the metal squares increases.  

A limiting factor of this design is the highest frequency at which these elements transmit.  This cut-off frequency is determined by a number of parameters, but most dominant is when the effective layer-to-layer thickness (i.e. substrate thickness) becomes less than half a wavelength.  Other effects occur due to field configurations at the corners of metal-meshes at higher frequencies, and give minimum gap sizes regardless of other geometric features.

\begin{figure}[ht!]
\includegraphics{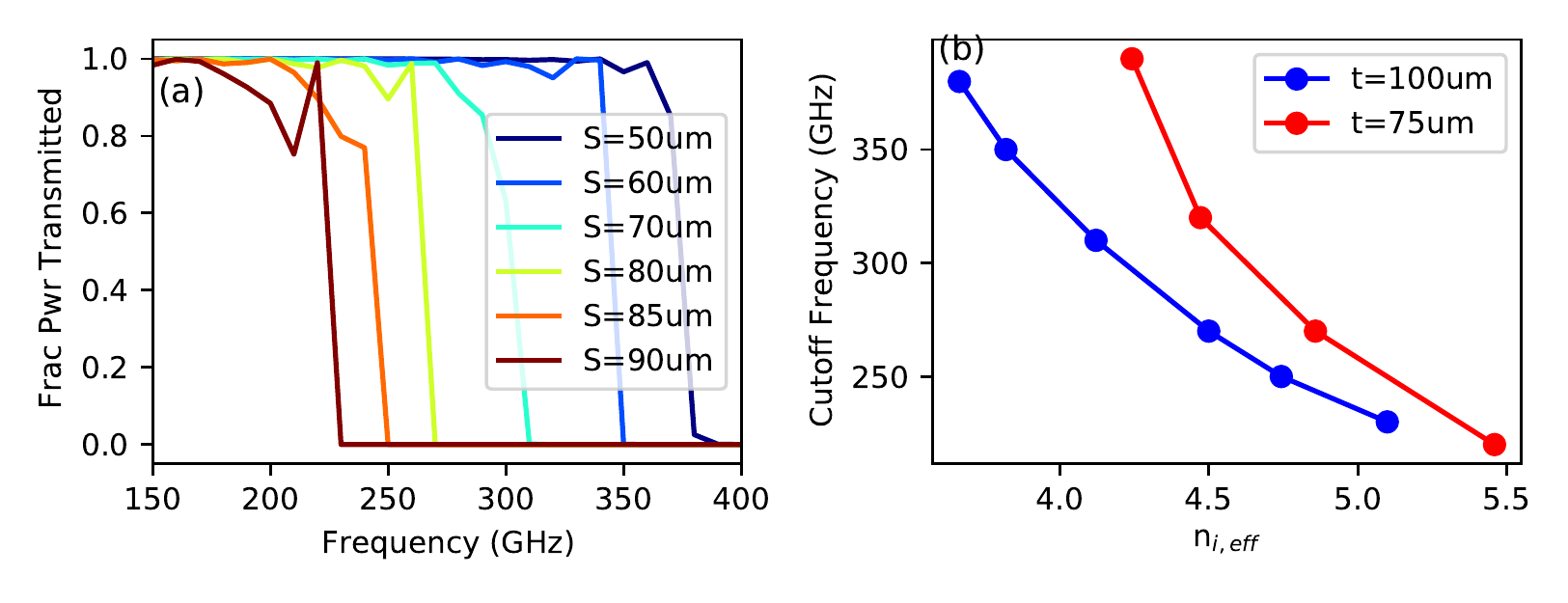}
\caption{(a) Forward transmission ($|S_{21}|^{2}$) for a metal-mesh metamaterial with a 100$\mu$m pitch and 100$\mu$m layer-to-layer spacing, using the geometry in Fig. \ref{fig:DielectricRot}(a) and assuming a silicon substrate ($\epsilon_{r} = 11.7$).  As the metal-mesh element size and thus also the effective index increases, the forward transmission is reduced at lower frequencies.  (b) The cutoff frequency as a function of effective index.  By going to thinner layer-to-layer spacing, higher frequencies can be achieved with the same effective index.}
\label{fig:MetalMeshCutoff}
\end{figure}

These results are shown in Fig. \ref{fig:MetalMeshCutoff}(a), where simulations similar to the geometry shown in Fig. \ref{fig:MetalMeshTop} were carried out over a broad frequency range.  The fractional power transmitted has a maximum frequency which decreases with side length.  

For a given substrate thickness this ultimately constrains either the peak effective index that can be used in designing a lens, which can affect GRIN performance, or sets the maximum frequency at which the lens can efficiently operate.  If higher frequencies or higher optical indices are required, thinner substrates will be necessary.  
The combined parameter space of index limit and maximum frequency is shown for two different silicon substrate thicknesses in Fig. \ref{fig:MetalMeshCutoff}(b).

\subsection{Etched holes in silicon}

Metamaterials can also be fabricated by etching sub-wavelength holes in a dielectric such as silicon, creating a material with an index varied by adjusting the ratio of free space and silicon.  The minimum index (maximum free space filling fraction) is constrained by fabrication limits -- enough silicon must be left to remain mechanically robust.  Additional constraints are imposed by uniformity of etch and desired feature resolution, although at mm-wavelengths the micron-scale uniformity provided by standard lithography processes is more than sufficient. 

\begin{figure}[h!]
\includegraphics{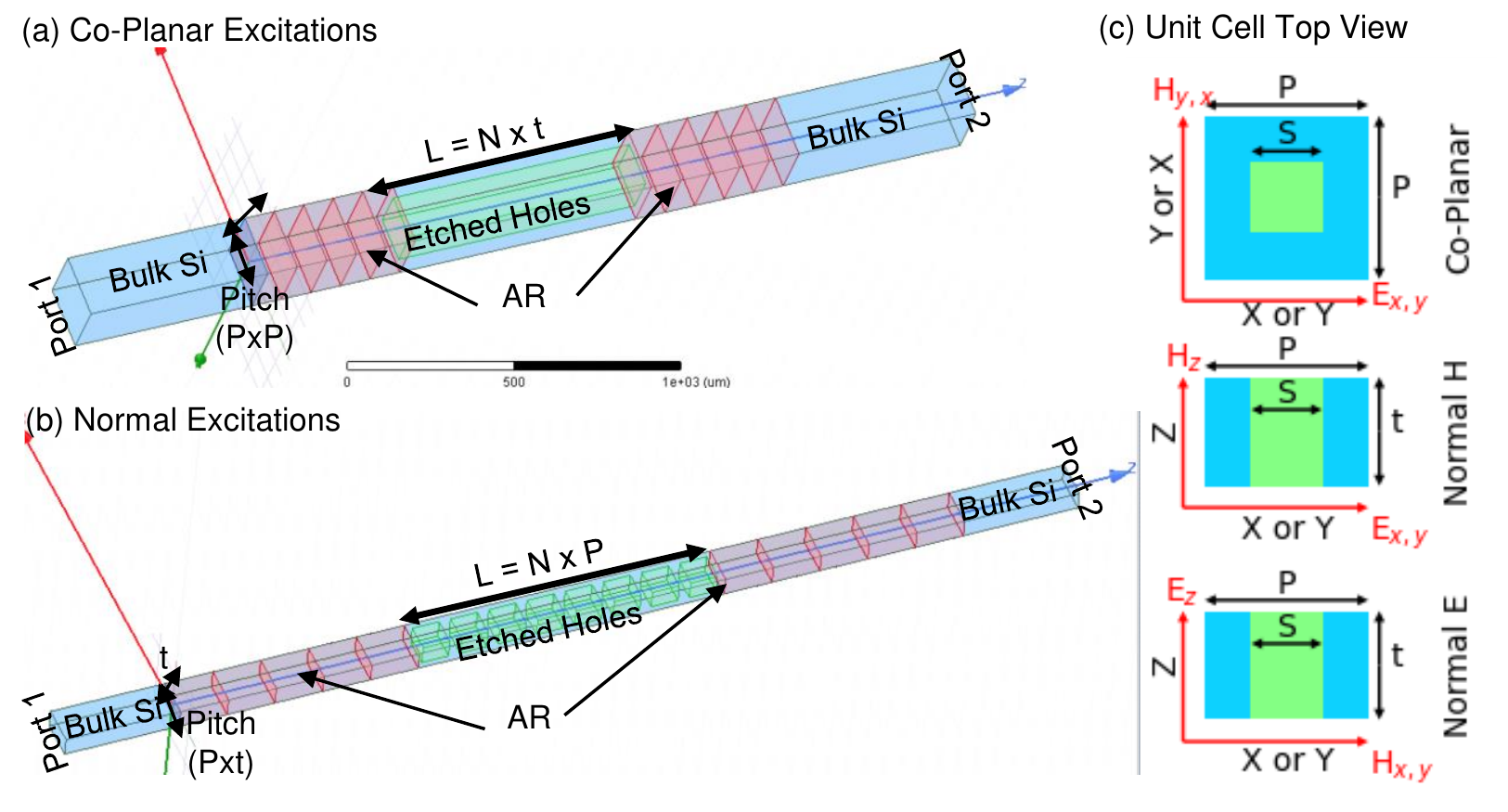}
\caption{Schematic of structures used to calculate equivalent optical parameters of sub-wavelength holes etched in silicon substrates.  (a) Plane wave excitation normal to the lens with the E and H fields Co-Planar to the lens.  The structure is excited from ports in bulk silicon.  Several intermediate index materials, labeled as AR, are used to minimize reflections in the simulation.  The length of the etched hole (green) is varied to calculate phase velocity.  (b) A similar structure, but now for the plane wave incident tangent to the lens.  The holes are no longer continuous in the z-axis.  (c) Top view of the unit cells for the three types of excitations.     }
\label{fig:DielectricRot}
\end{figure}

The unit cells used for simulation are shown in Fig \ref{fig:DielectricRot}, with similar parameters to those described previously for metal mesh elements.  Fig. \ref{fig:DielectricRot}(a) shows a stack of unit cells, which are realized as continuous holes along the z axis when the substrates are stacked.  Fig. \ref{fig:DielectricRot}(b) shows the hole geometry for simulations of a plane wave traveling tangent to the detector plane.  The three types of incidence are as described in the previous section, and the top view of each unit cell is shown in Fig. \ref{fig:DielectricRot}(c).

As there are no conductors in this metamaterial, all permeabilities are equal to that of free space ($\mu_{eff,Co} = \mu_{eff,z} = 1$) and Eq. \ref{Eq:phasevel} reduces to
\begin{subequations}
\begin{align}
&v_{ph,Co} = c / \sqrt{\epsilon_{eff,Co} } \\
&v_{ph,Hz} = c / \sqrt{\epsilon_{eff,Co} } \\
&v_{ph,Ez} = c / \sqrt{\epsilon_{eff,z} } 
\end{align}
\label{Eq:phaseveldiel}
\end{subequations}
where $c$ is again the speed of light in free space and $\epsilon_{eff,Co}$ and $\epsilon_{eff,z}$ are the permittivities along the different directional components.  

Achieving lower indices requires a larger pitch to etch sufficient silicon while preserving a mechanically robust structure.  However, the maximum pitch is constrained by the maximum frequency and equivalent index desired.  In general a metamaterial must keep its pitch to less than half an effective wavelength in media to avoid scattering to other modes, but as the frequency increases the microscopic field properties can cause the effective index value to diverge from the low frequency limit even as the material continues to operate as a metamaterial.  Divergence from the desired value can have unwanted impact on lens performance that varies with frequency.

\begin{figure}[h!]
\includegraphics{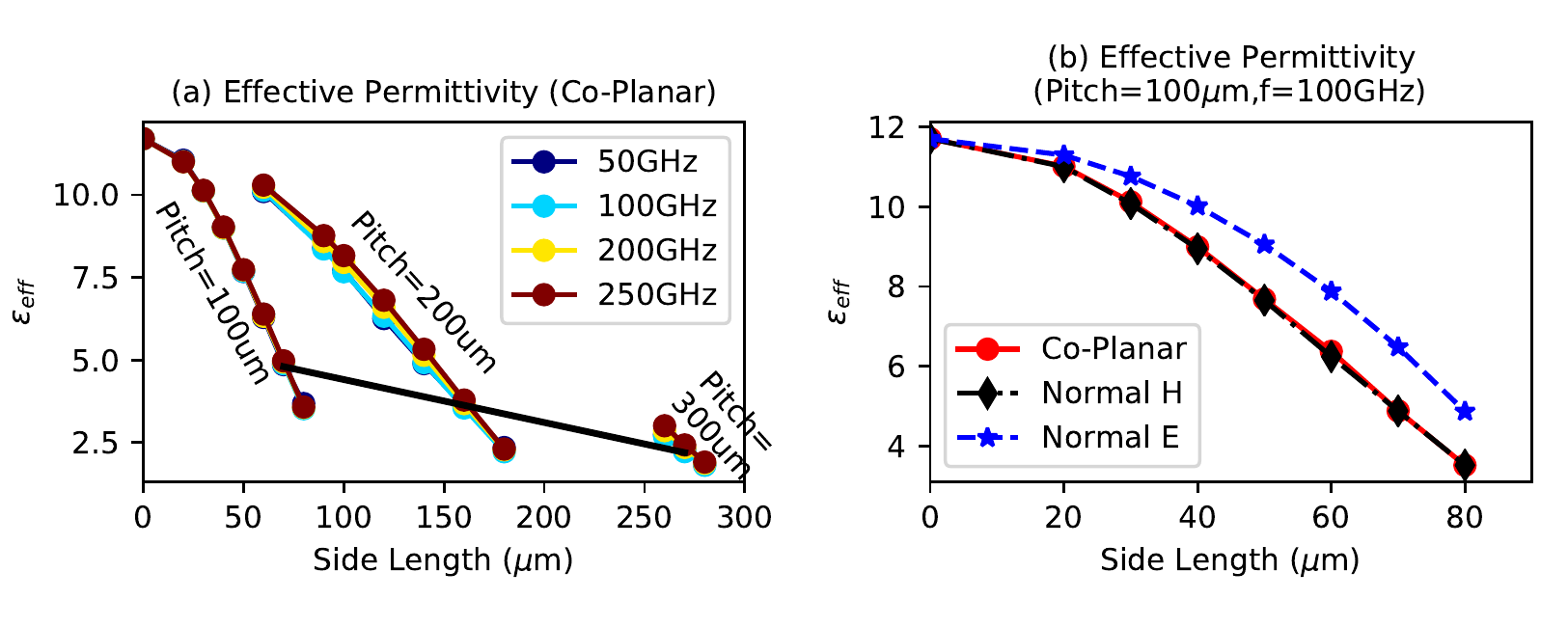}
\caption{(a) Effective permittivity for various etched-hole side lengths and pitches along the primary Co-Planar excitation.  The three sets of colored curves are for three different pitch sizes (distance between hole centers).  The solid black line is the minimum achievable permittivity for a limit of a 30 $\mu$m thick silicon wall for each pitch.  For the two larger pitches, the higher frequency index diverges at higher indices, and therefore the top point represents the highest achievable permittivity using that pitch while exhibiting less than 10\% variation in effective index and less than 1\% scattering to other modes.  (b) The fit equivalent parameters for all three excitations.  Permittivity is fit using the phase velocity.  As expected from Eq.~\ref{Eq:phaseveldiel}, permittivities for the Co-Planar and Normal H excitations have identical values.  Higher effective permittivities are calculated for the Normal E excitation. }
\label{fig:MultiPitchIndex}
\end{figure}

This trade off is shown in Fig. \ref{fig:MultiPitchIndex}(a).  Simulations with different hole sizes are carried out on different pitches and results are shown for the Co-Planar excitation.  As expected, larger holes (larger free space filling fractions) result in lower effective permittivity.  However, for any given pitch there is a maximum frequency of operation, shown as the highest point on each plot where power is still efficiently transmitted through the metamaterial at 250 GHz.  As the pitch increases this naturally decreases relative to the increase in effective wavelength.  A solid black line is shown in Fig.~\ref{fig:MultiPitchIndex}(a) and represents the minimum effective permittivity that can be constructed with a minimum silicon free standing wall thickness of 30 $\mu$m, a thickness we have fabricated and demonstrated to be mechanically robust.

Given these constraints, an efficient strategy to use these materials would include using several pitches within the same structure based on required frequency and optical indices.  Such an approach would allow Co-Planar excitations to access effective permittivities from the bulk ($\epsilon_{r,Si} \approx 11.7$) to values $\epsilon_{Co,eff} \sim 2$.  

The frequency behavior from 50 to 250 GHz for this metamaterial is shown for all three excitations in Fig. \ref{fig:MultiPitchIndex}(a).  At larger pitches the effective permittivity at higher frequencies begins to diverge from the low frequency value, even as power still efficiently transmits.  At the smallest pitch ($100 \; \mu\text{m}$) the variations are sub-percent; however, at the larger pitches variations of several percent become apparent.  While this will not necessarily significantly degrade transmission, it will result in frequency-dependent performance that must be considered in modeling these materials at higher frequencies.

The results of exciting these structures with the \textit{Normal E} and \textit{Normal H} excitations are shown in Fig. \ref{fig:MultiPitchIndex}(b) for a pitch of 100 $\mu$m and a frequency of 100 GHz.  While the \textit{Co-Planar} and \textit{Normal H} show identical results as expected, the \textit{Normal E} geometry has a very different corner structure, which results in effectively higher permittivity.  However, it is worth noting that the intended use for these materials is in anti-reflection coatings, where the radiation should be well collimated and dominated by the co-planar field conditions.  In applications where this is not true, additional simulations may be required. 


\section{Prototype Optical Lenslet}

A prototype lenslet was developed based on the metal-mesh metamaterials described.  This lenslet mates to an existing PB/SA PB2a focal plane array designed for operation at 90 and 150 GHz, with pixels in a hexagonal packing scheme with a pixel pitch of 6.7 mm.  An existing seating wafer separates the lenslet array from the detector through a total silicon thickness of 1.1 mm.  The seating wafer is aligned by the PB/SA collaboration with a minimum 25 $\mu$m accuracy to the antenna.

\subsection{Initial Prototype Design \& Simulations}

The GRIN geometry alters the effective phase velocities such that the incident ray from the antenna is collimated at the exit of the structure.  While complex interactions within near field of the antenna and lens affect the required design, as well as realized optical paths, an initial seed geometry design is used and then iteratively refined.

To develop a seed geometry, the sinuous antenna is first simulated using an infinite half-space of silicon.  Phases at the wavefront are calculated and initial geometry chosen using the simple equal-path estimation of Eq.~\ref{eq:length}, i.e. using the center index ($n_{1,eff}$) and edge index ($n_{7,eff}$) as well as the phase difference between the center and edge at for a given lens focus, as shown schematically in Fig. \ref{fig:GRINBasics}.  After traveling the length of an optimally designed lens, the phases are equal at the exit.  For this particular antenna, simple calculations and simulations suggested a lens focus of 2.0 mm and length of 3.4 mm.

For this prototype lenslet simulations were initially carried out using a symmetric quarter-model geometry with fully-modeled metal mesh elements.  This approach was chosen as it reduced the memory requirements to something our 1 TB server could solve.  Quarter-model symmetry is exploited using a twin slot antenna with geometry modified to approximate sinuous antenna behavior at a particular frequency.  Metal meshes are drawn on a 125 $\mu$m grid and the overall geometry split into seven radial regions, with identical mesh sizes in each region.  Each region is bounded on top and bottom by metal mesh squares that taper from the GRIN metal mesh size to just bulk material, serving as an AR region between the high index material and bulk silicon.

\begin{figure}[ht!]
\includegraphics{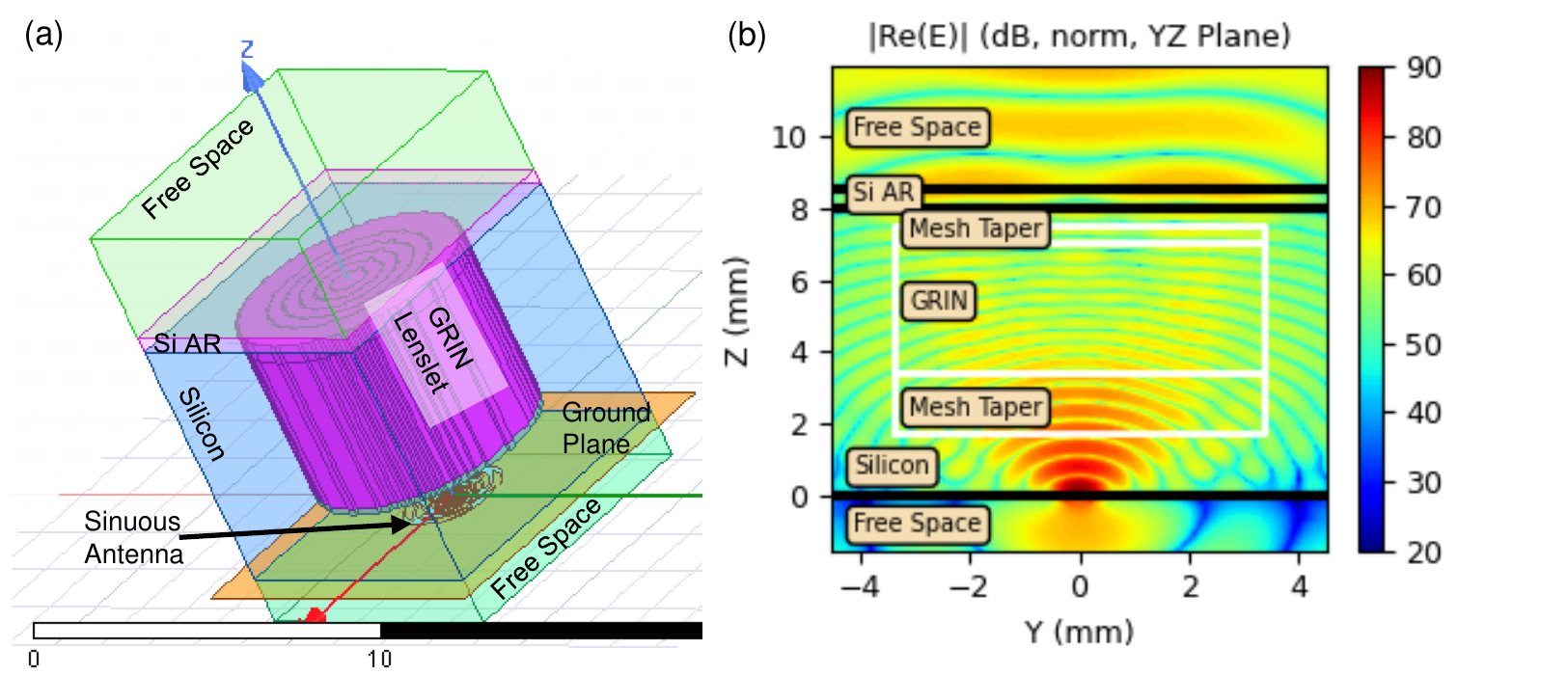}
\caption{(a) HFSS schematic of the GRIN lenslet.  A sinuous antenna is defined on perfectly conducting ground plane (orange).  The antenna is coupled to the GRIN lens through the bulk silicon (blue).  The various GRIN layers and coupling components are shown (purple).  Etched holes in silicon forming an AR coating (pink) to free space (green) are shown.  The backside of the antenna couples to free space (green), but the higher index silicon allows $\sim 95\%$ of the power to couple in the direction of the silicon substrate.  (b) Cross section of the electric field from an HFSS simulation, overlaid with description of lens components.   }
\label{fig:SGRinDesign}
\end{figure}

The overall length of the lens, length of the tapers, mesh sizes in the seven radial regions, and lens focus were treated as adjustable parameters and simulations run to optimize optical efficiency within the Lyot stop for the PB2/SA optics (+/- 14$^{\circ}$).  Parameters were varied until the model converged to an optimized geometry, with overall geometry shown schematically over simulated fields in Fig. \ref{fig:SGRinDesign}(b).  

\begin{figure}[ht!]
\includegraphics{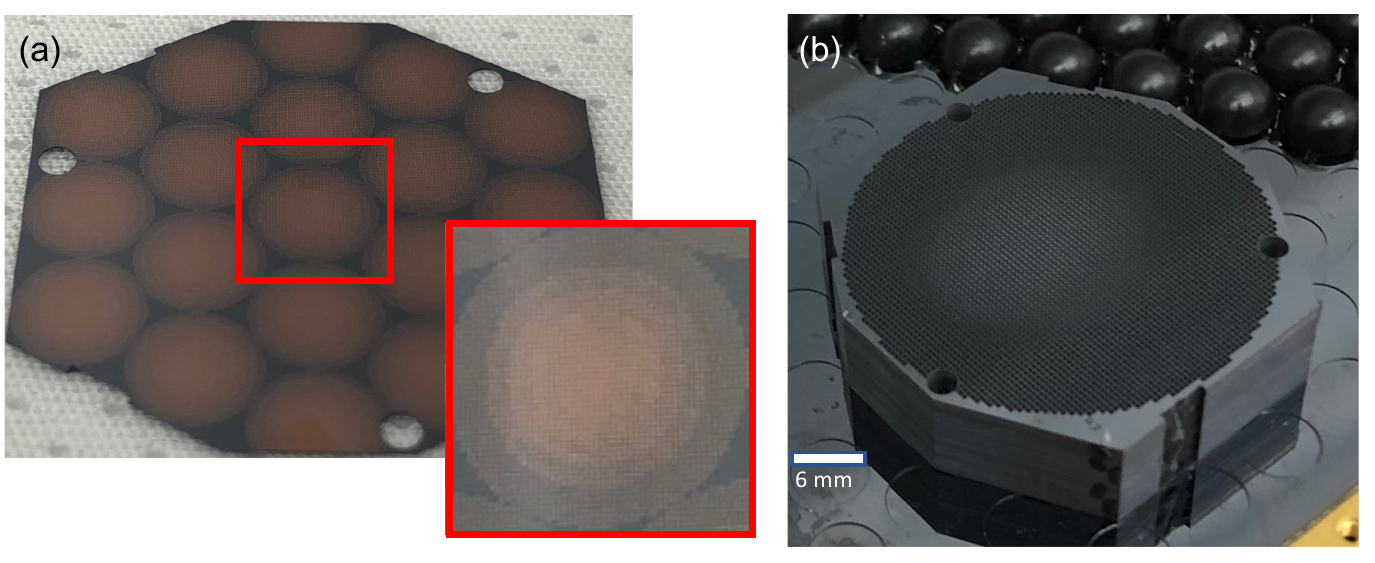}
\caption{Lenslet construction.  (a) A single 100$\mu$m thick silicon platelet with metal mesh squares.  Also visible are holes for clamping and cut outs along the edges for alignment and bonding the stack with epoxy. (b) The full 19 pixel array mounted to an existing PB-2 focal plane.  Also visible in the background are hemipsherical lenslets, individually glued to the interface seating wafer.  The top of the lens structure is the AR layer made from holes etched in silicon.}
\label{fig:Build}
\end{figure}

The optimized GRIN lenslet comprises 36 identical layers, each 100 $\mu$m thick for a total length of 3.6 mm, which can be seen as-built in Fig.~\ref{fig:Build}(a), where a single metalized substrate is shown.  The center region has squares with side length 97.5 $\mu$m, with side lengths stepping down to 55 $\mu$m at the edge.  This was mated to additional elements which tapered from the designed side length to free space on either size of the lens, for a total of 53 layers and length of 5.3 mm.  

Silicon substrates were also etched to serve as various spacer wafers between the antenna and lenslet and between the GRIN lens and the free space AR section.  Etched holes in silicon as described in Sec. \ref{sec:construction} were used.  We used on-hand silicon wafers which were 250 $\mu$m thick and a two-layer AR geometry as shown in Fig. \ref{fig:ARBuild}(a).  On the higher index side of the two-layer structure, 70 $\mu$m square holes are etched on a 162.5 $\mu$m pitch.  On the lower index side a 260 $\mu$m hole is etched on a 325 $\mu$m grid. Simulations suggest this will provide better than 90\% optical efficiency at the frequencies of interest.  While better AR coatings could be designed with more or thinner layers, this thickness was chosen only because it was a standard thickness we had stocked.

\begin{figure}[ht!]
\includegraphics{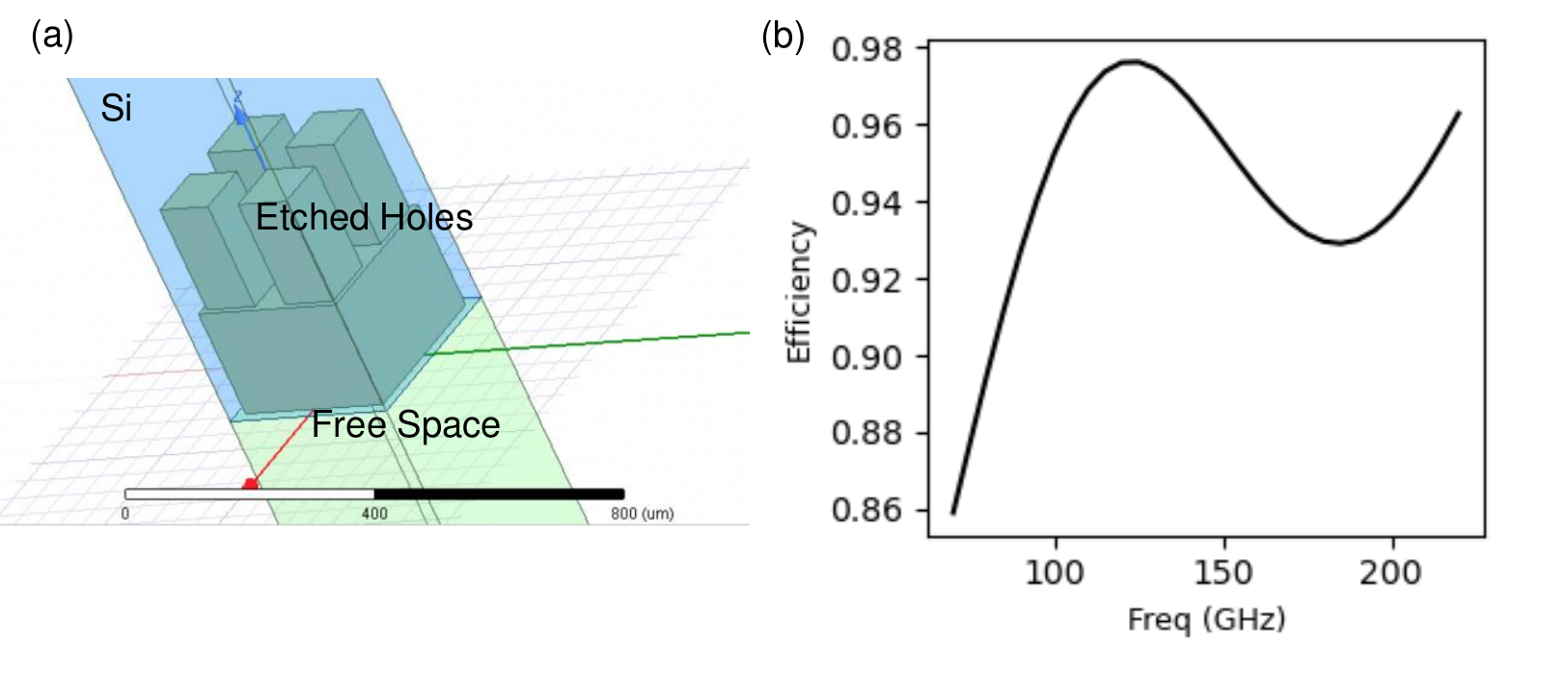}
\caption{Anti-reflection layers for the prototype lenslet array.  (a) HFSS schematic of a single cell simulated in an infinite periodic array.  The layer consists of two silicon substrates, each 250 $\mu$m thick, with etched hole patterns and stacked.  (b) Simulations of the structure showing performance over the 90/150 GHz bands. }
\label{fig:ARBuild}
\end{figure}

The bulk-model discussed in Sec. \ref{sec:construction} had not been fully developed at the time this lenslet was designed.  The design process was optimized to couple radiation from the twin slot antenna in to the GRIN lens, and sensitive to the near-field characteristics of the antenna.  A single, full-wave simulation was run on a machine with 2 TB of memory, with sinuous antenna and full metal mesh elements, and at the time we believed it had fully converged.  However, in developing the bulk-model and investigating far-field convergence, we discovered this was not the case, as the convergence criteria were not clear, as discussed in \ref{sec:Memory Usage}.

We therefore re-simulated the lens with our bulk-model equivalent in parallel with the fabrication approach.  A cross section of the fields from this simulation is shown in Fig.~\ref{fig:SGRinDesign}(b).  One weakness discovered is that the near-field of the sinuous antenna was was sufficiently different from the twin slot to cause about 25\% of the power to radiate in to the silicon due to poor optical coupling.

Poor coupling has two deleterious effects on this lenslet.  First, it has an overall negative impact on the optical efficiency.  Secondly, in trying to understand our models and compare to measurements, a significant amount of power will be reflecting around the silicon with unknown impact on the beam patterns.  However, we decided that these effects were small enough to not warrant re-design, and to measure the lenslet as is.

\subsection{Lenslet Array Construction \& Measurements}

The lenslet is assembled by first stacking the various 100$\mu$m thick silicon wafers with the lithographed metal mesh patterns.  
In addition, the anti-reflection structure made using etched holes in silicon from a stack of two 250 $\mu$m thick wafers with different hole patterns, as described in Fig. \ref{fig:ARBuild}, is placed on top.  Finally, a seating wafer is made by using a DRIE approach to create bosses which mount to the seating pockets visible in Fig. \ref{fig:Build}(b).

All of these layers are stacked.  DRIE etched holes are used for screws which temporarily tighten the assembly together during construction while granite blocks are used to align corner features to achieve better than $10\mu\text{m}$ layer-to-layer alignment.  The full assembly is epoxied together with a thin bead of Stycast 2850, which is robust to thermal cycling, and the screws are removed.

The overall geometry is a hex-packed array of 19 lenslets.  The lenslet stack is mounted to a seating wafer with bosses, using lithographically-defined mating features on the lenslet wafer which allow for simple mounting with alignment accuracy.  Rubber cement is then used to secure the detector array to the focal plane.

\begin{figure}[h!]
\centering
\includegraphics[width=0.95 \textwidth]{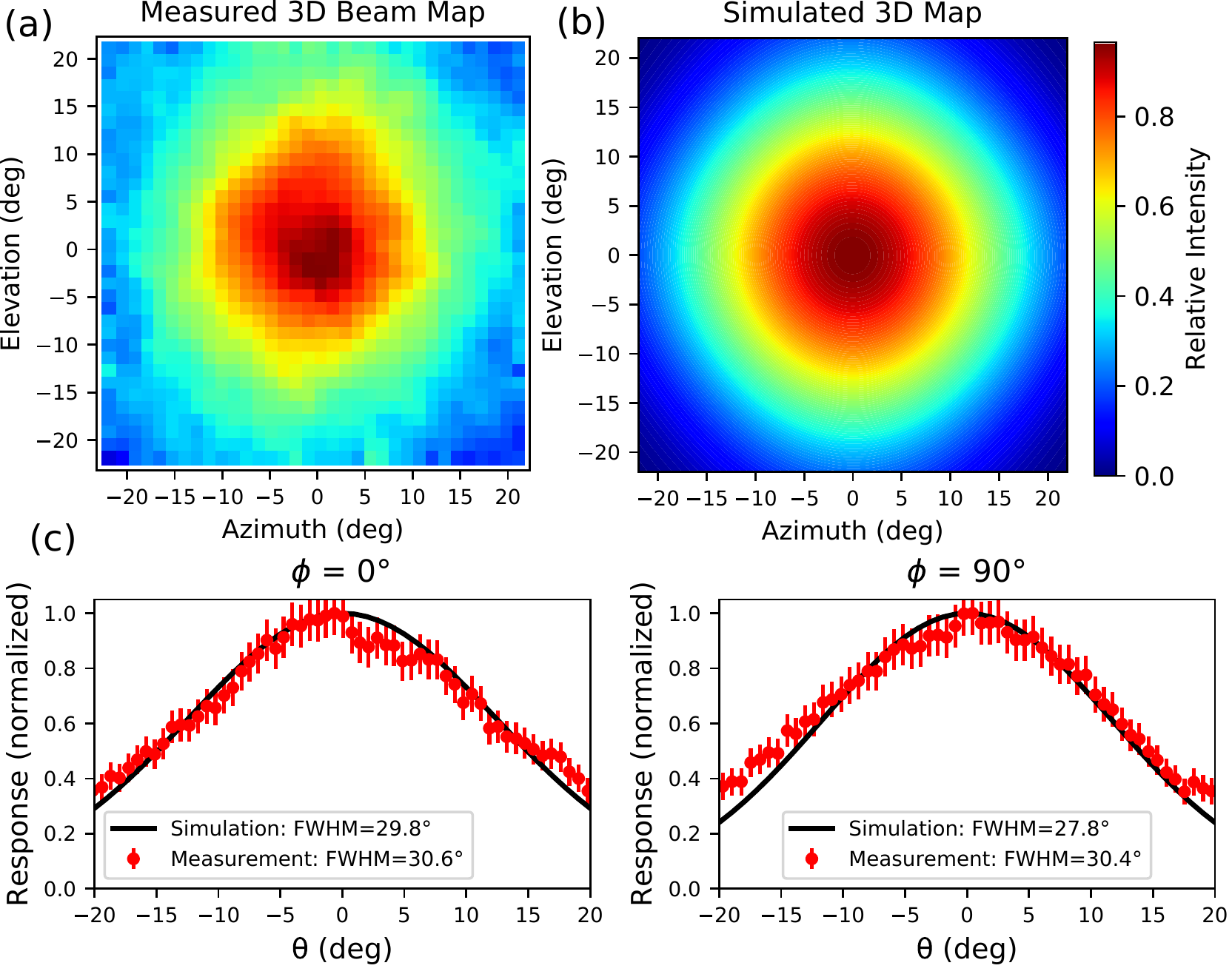}
\caption{Si GRIN Lenslet maps.  (a) Pointed 3D beammap of the GRIN-lenslet read out with the 90 GHz band of a PB/SA PB2a focal plane and illuminated with a broadband thermal source. (b) Single frequency (band-center) simulation results from HFSS.  (c) Plots of the measured broadband 90 GHz response along $\phi = 0^\circ$ (Azimuth = $0^\circ$) and $\phi = 90^\circ$ (Elevation = $0^\circ$) and comparison to simulation data. }
\label{fig:SiGRINMaps}
\end{figure}

We show in Fig.~\ref{fig:SiGRINMaps} the results of measurement compared to simulation.  For this measurement we used a spare PB/SA PB2a detector wafer which had lower than desired yield and slightly offset ($\sim 8 GHz$) frequency bands.  The detector pixel under the center of our 19-pixel prototype lenslet array was not functional.  Therefore, we measured a lenslet which was not the center of the array.  Beam maps were done using a thermal source mounted on a six degree of freedom (6DOF) beam mapper, which allows the source to point directly at our detector, giving a true 3D beam map.  

The results of the measured beam map are shown in Fig.~\ref{fig:SiGRINMaps}(a), and  simulated response in Fig.~\ref{fig:SiGRINMaps}(b).  While the overall shapes are roughly similar, we note the measured map shows more response at wider angles, as well as a slightly elongated shape.  The higher response may be due to a number of sources, such as the stray light leaking in to the silicon substrate described above, or even optical effects as we approach the clipping angles of our 3D beam mapping system, which has hard cut-offs above 25$^{\circ}$.  These are all currently under investigation.  Regardless, the overall similarity as well as the 1-D cuts shown in Fig. \ref{fig:SiGRINMaps}(c) give sufficient confidence in comparing simulation to measurements.

While the absolute optical efficiency is difficult to directly measure given many uncertainties in the system, we did measure the relative optical response of the GRIN lenslet and existing hemispherical lenslet, measured side-by-side on the same array as shown assembled in Fig. \ref{fig:Build}(b).   The measured response ratio showed the GRIN lenslet with 65\% of the response compared to the hemispherical lenslet, very similar to the simulated ratio of 68\%.  In simulation this lower optical efficiency is almost entirely accounted for by the power leaked in the silicon from the poor coupling design, and not any inherent property of a GRIN lens.

\clearpage

\section{Conclusions and Future Work}

The GRIN lenslet tested above was designed to mate to an existing detector focal plane.  This set a minimum distance between detector and lenslet, which in a detector wafer designed inherently for GRIN compatibility could be as small as the detector wafer thickness.  Preliminary simulations of such geometry where the lens is moved within 500 $\mu$m of the antenna suggest that the spillover efficiency increases from $\sim$ 35\% for a traditional hemispherical lenslet at 90 GHz to to 45\% with an optimized GRIN, suggesting mapping speed improvements of at least 25\% are possible.

For existing detector wafers and testing we are limited by the minimum 1.1 mm distance.  Despite this, we have designed a lenslet which addresses the sinuous to lenslet coupling issues encountered in the prototype lenslet. Simulation suggests that even with a mounting scheme optimized for hemispherical lenslets we can achieve equivalent spillover efficiency and validate our understanding of sinuous antenna interactions with the GRIN lenslet.

The planar GRIN also offers a simple approach to mounting silicon-based planar AR structures.  As shown in Fig. \ref{fig:DielectricRot}(b), holes etched in silicon allow a broad range of optical indices to be accessed, and silicon wafers, especially silicon on insulator (SOI), allow for many custom thicknesses to be fabricated.  Thus an adiabatic AR structure can be developed which transitions slowly from the bulk silicon to free space.  Double sided SOI wafers with thicknesses of 50 $\mu$m are feasible for fabrication, allowing for operation at frequencies even into the sub-mm regime. 

Simulations have been carried out to explore the feasibility of broadband AR coatings. A general design philosophy is to divide a structure into sub-structures which are less than a quarter effective wavelength long at the highest frequency of operation, with an overall length of more than half a wavelength at the lowest frequency. For the lowest indices before free space, we have fabricated suspended metal mesh structures on SiN substrates, and the silicon wafers they are suspended on have been fabricated with thicknesses as low as 125 $\mu$m.  

\begin{figure}[ht!]
\includegraphics{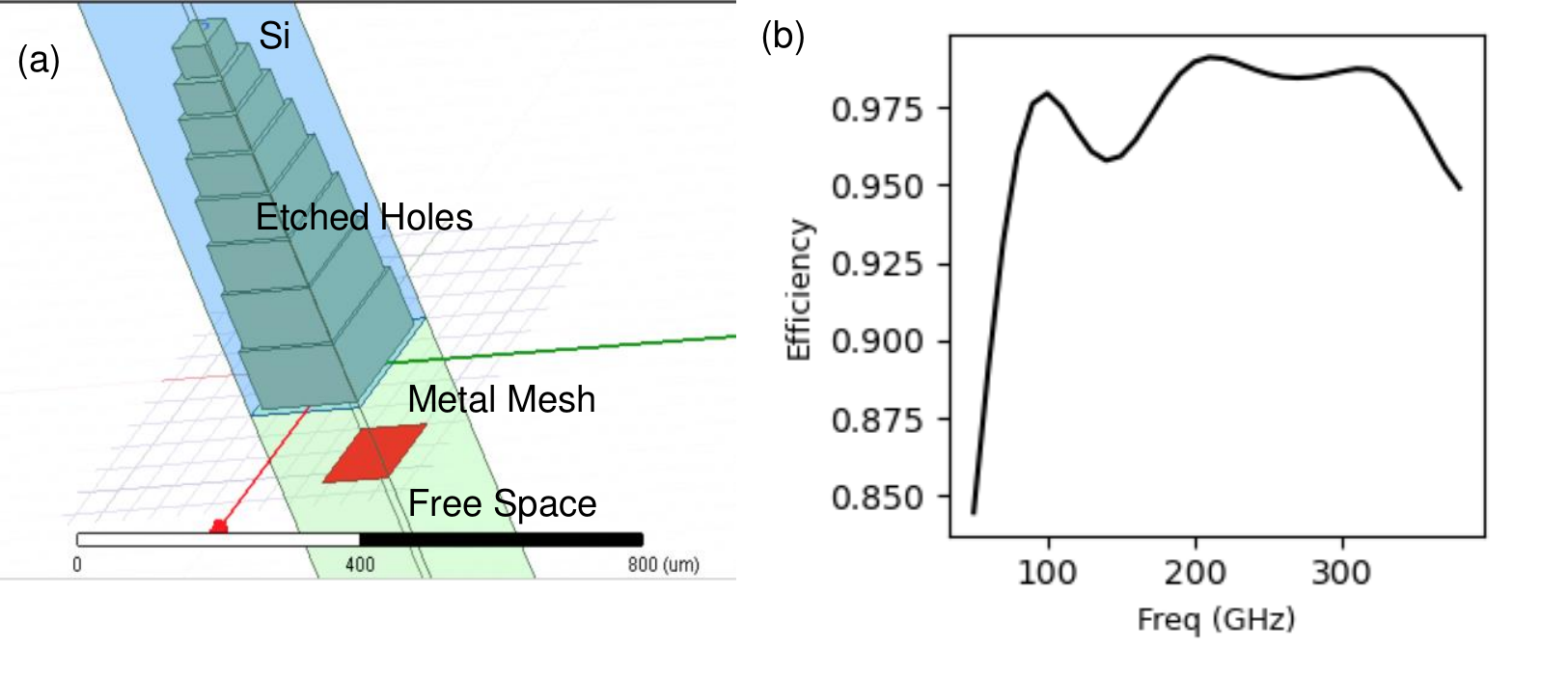}
\caption{Anti-reflection layers proposed. (a) HFSS schematic of an adiabatic AR structure.  This is a single unit cell in a periodic array.  The hole sizes and lengths increase as the structure moves from silicon to free space.  A final layer is synthesized using a metal-mesh element suspended on a SiN substrate to achieve a low index.  The design philosophy is discussed in text. (b) Simulations of the structure, showing better than 95\% efficiency from 70 to 350 GHz.}
\label{fig:ARBuild}
\end{figure}

Simulation results of this structure are promising, with better than 95\% efficiency over a range from 70 GHz to 350 GHz, with a total thickness of 625 $\mu$m.  This total thickness is not considerably thicker than the 580 $\mu$m required for a quarter-wavelength, index matched AR coating for use at 70 GHz. 

We have presented a planar, lenslet geometry which is compatible with broadband AR layers.  Measurements closely match simulations which have been improved to allow an iterative design cycle and physical interpretation of the metamaterial elements. This technology offers advantages over traditional coupling schemes in uniformity, yield and mapping speed.  Additionally, the natural ability to couple to planar broadband AR structures and broadband sinuous antennas without complex backshorts allows considerable scaling in frequency space by varying lithographic designs.  The ability to fabricate these structures with high yield and monolithically over an entire wafer makes them valuable for large-scale arrays needed for next generation instrumentation.

\appendix    

\section{Memory Usage and Bulk-Model Equivalence}
\label{sec:Memory Usage}

To study the efficacy of our metamaterial equivalent solutions over a broad range of frequencies, we created a quarter-symmetry model using a twin slot antenna and feeding a lens structure.  Equivalent lens structure were modeled using bulk materials and metal-mesh grids.  The geometry was chosen such that it could converge and continuing iterating for many cycles using cluster computing resources with 4 TB of memory.

\begin{figure}[ht!]
\includegraphics[width=0.95 \textwidth]{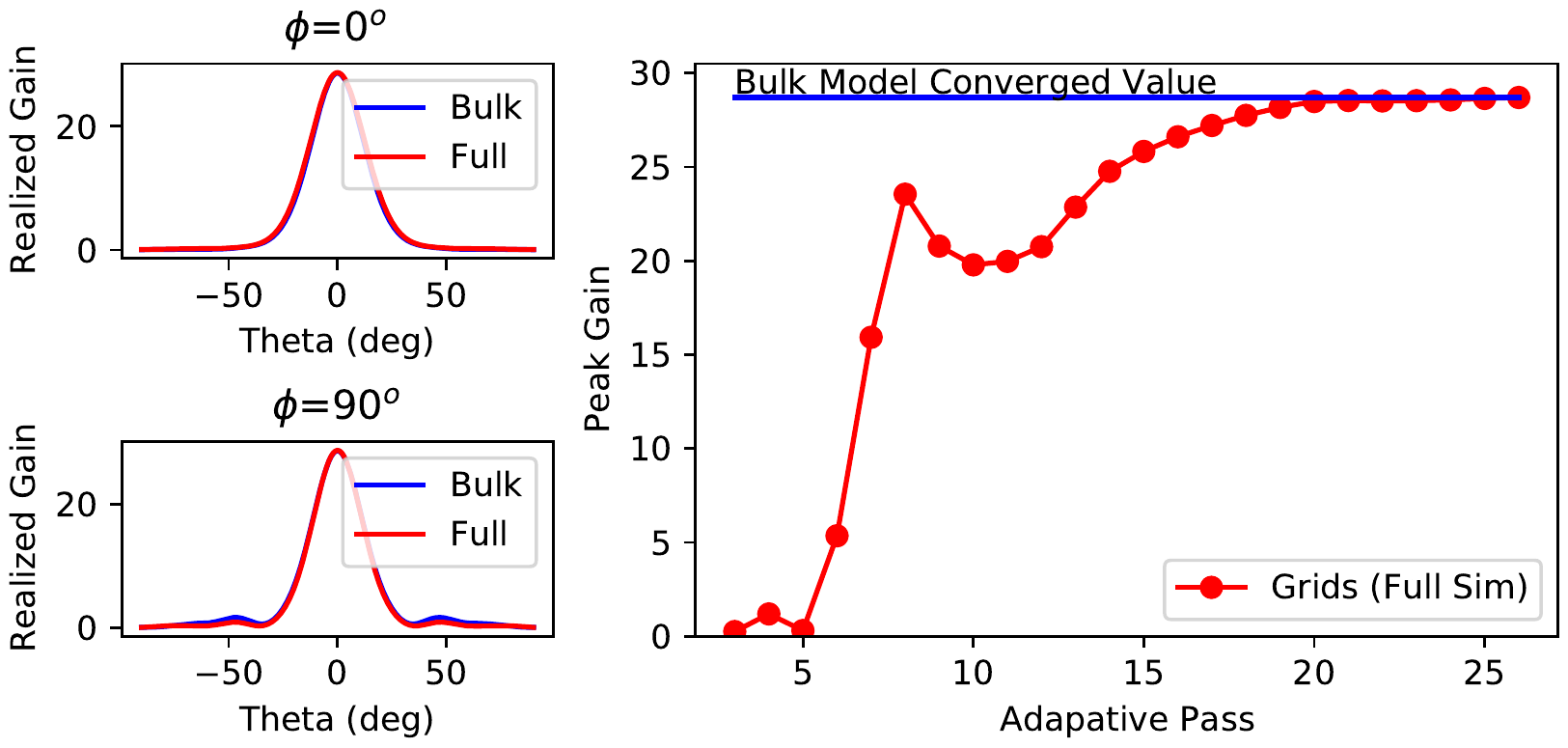}
\caption{Simulations of a lens comparing bulk equivalent model to full metal mesh grids.  Simulations carried out on a cluster computing resource with 8 TB of memory available. The left panel shows an overlay of the far field results from both models.  The right panel shows the convergence of the peak gain as a function of iteration number.  HFSS reported the simulation as converged at pass 12, with 600 GB of memory. However, the far field clearly continues to evolve and does not really converge until about 20 iterations, which required 2 TB of memory. }
\label{fig:CompareConvergence}
\end{figure}

As shown in Fig. \ref{fig:CompareConvergence}, the solution required a significant number of iterations to fully converge.  Deceptively the FEM modeling software, which uses the ports to evaluate convergence, suggested it had converged around iteration 12, at only around 600 GB of memory.  However using far-field results as the metric of convergence, such as the peak gain shown, revealed the structure had not truly fully converged until around 20 iterations and 2 TB of memory.

The 2 TB memory usage is consistent with simulations of individual elements memory requirements to converge effective optical parameters to within 2\% of their desired values.  Rough calculations suggested that for lens lengths of 3 mm and the indices used this should not alter the phase front from its fully converged value by more than 5\% along any path, which would not disturb far field calculations considerably.

Fig~\ref{fig:CompareConvergence} indicates agreement between the bulk-model, which required 100 GB and only a few hours to run on a 12-core machine, and the full FEM simulation, which required nearly 6 days on a machine with 4 TB of memory available.  Along $\phi=90 ^{\circ}$ there exists a small discrepancy in the first side lobe.  This might be due to difficulties in fully capturing physics at the interface especially along the H-plane, where the bulk-model had more difficulty developing proper fits.  Regardless, these variations are small and justified for savings in time and memory.

This suggests a successful process for lenslet design.  The lenslet parameters can be adjusted using the bulk material models.  With modern cluster computing resources a bulk model can be run at 90 GHz in a few hours and 150 GHz in 12 hours. Therefore a lenslet can be optimized in a few week time scale.  A final design can then be fully run using the 8 TB of memory, which will take of order 1-2 weeks and result in a simulation that can be compared to measurement and bulk-model simulation.  While generally very close to the bulk model equivalent, this final simulation may illuminate unexpected behavior or differences in side lobes and is therefore worthwhile.

\acknowledgments 
The material in this paper is based upon work supported by NASA under grant/cooperative agreement number NNX17AE85G. This work utilized resources from the University of Colorado Boulder Research Computing Group, which is supported by the National Science Foundation (awards ACI-1532235 and ACI-1532236), the University of Colorado Boulder, and Colorado State University. The lenslets were fabricated in the NIST Boulder Microfabrication Facility (BMF).

\bibliography{meshlens} 
\bibliographystyle{spiebib} 

\end{document}